\documentclass[aps,prd,superscriptaddress,nofootinbib,amsmath,amsfonts,preprintnumbers,groupedaddress,showpacs,10pt,english]{revtex4-1}
\usepackage{amsmath}
\usepackage{amssymb}
\usepackage{babel}
\usepackage{wrapfig}
\usepackage{cancel}

\usepackage{relsize,exscale}
\makeatletter

\usepackage{array,multirow,graphicx}
\usepackage{dcolumn}
\usepackage{newlfont}
\usepackage{bm}
\usepackage[colorlinks,citecolor=blue,urlcolor=blue,linkcolor=blue]{hyperref}
\usepackage[figtopcap]{subfigure}
\usepackage{color}

\begin{document}

\date{\today}

\title{Nonlinear charged black hole solution in Rastall gravity}

\author{G.G.L. Nashed}
\email{nashed@bue.edu.eg}
\affiliation {Centre for Theoretical Physics, The British University, P.O. Box
43, El Sherouk City, Cairo 11837, Egypt}
\begin{abstract}
We show that the spherically symmetric { black hole (BH)} solution of a charged (linear case) field equation of Rastall gravitational theory is not affected by the Rastall parameter and this is consistent with the results presented in the literature. However, when we apply the field equation of Rastall's theory to a special form of nonlinear electrodynamics (NED) source, we derive a novel spherically symmetric BH solution that involves the Rastall parameter. The main source of the appearance of this parameter is the trace part of the NED source, which has a non-vanishing value, unlike the linear charged field equation. We show that the new BH solution is just Anti-de-Sitter Reissner-Nordstr\"om spacetime in which the Rastall parameter is absorbed into the cosmological constant.  This solution coincides with Reissner-Nordstr\"om solution in the GR limit, i.e. when Rastall's parameter vanishing. To gain more insight into this BH, we study the stability using the deviation of geodesic equations to derive the stability condition.  Moreover, we explain the thermodynamical properties of this BH and show that it is stable, unlike the linear charged case that has a second-order phase transition. Finally, we prove the validity of the first law of thermodynamics.

%\pacs{ 04.50.Kd, 98.80.-k, 04.80.Cc, 95.10.Ce, 96.30.-t}
\end{abstract}
\maketitle
%%%%%%%%%%%%%%%%%%%%%%%%%%%%%%%%%%% Section 1 %%%%%%%%%%%%%%%%%%%%%%%%%%%%%%%%%%%%%%%%
\section{Introduction}\label{S1}
%%%%%%%%%%%%%%%%%%%%%%%%%%%%%%%%%%%%%%%%%%%%%%%%%%%%%%%%%%%%%%%%%%%%%%%%%%%%%%%%%%%%%%
Since the construction of Einstein's general relativity (GR), the coupling between a scalar field and the gravitational action in a geometric frame has been intensively studied. A scalar theory formulation was made in  \cite{nordstrom1912relativitatsprinzip}, and Jordan-Brans-Dicke later built a gravitational theory as an expansion of GR to investigate the variable of gravitational coupling \cite{schucking1999jordan,PhysRev.124.925,Dirac:1937ti}. Afterward, a general combination between a scalar field and its derivative, which yields second-order differential equations, is known as the Horndeski theory \cite{Deffayet:2013lga} that gained much attention. Recently, many modifications of Einstein GR have been established. Among these theories is the $f(R)$ gravitational theory, which is regarded as a natural generalization of Einstein's Hilbert action \cite{DeFelice:2010aj,Nojiri:2010wj,Nojiri:2017ncd}. This theory could be rewritten as a GR and scalar field \cite{Elizalde:2020icc,Nashed:2019yto}. The above is a very brief summary related to the scalar fields in the frame of a gravitational context. However, there is a huge literature on this subject.

The above discussions show one way of modification of GR. However, there is another possibility that has been used to generalize the kinetic term of the scalar field that is minimally coupled to the Einstein-Hilbert action. This possibility is called the k-essence theory \cite{ArmendarizPicon:2000ah}. This theory is used as an option to the usual inflationary models that use a self-interacting scalar field
\cite{ArmendarizPicon:2000ah,Nashed:2018efg,Nashed:2018cth,Nashed:2020kjh,ArmendarizPicon:1999rj,ArmendarizPicon:2000dh}. Recently, vacuum static spherically symmetric solutions have been derived for the k-essence theories \cite{ArmendarizPicon:2000dh}. Some novel patterns have been derived that involves a study of the event horizon. Nevertheless, interpolating such solutions as black holes was difficult because it is impossible to define a distant region from the horizon. Using the no-go theorem, it has been affirmed that solutions with a regular horizon can exist but only of the type of cold black hole  \cite{Bronnikov:1998gf, Bronnikov:1998hm}.

Another generalization of GR is to abound the restriction of the conservation law encoded in the zero divergence of the energy-momentum tensor. Among the theories that follow this direction is the one given by Rastall (1972), which is known as Rastall's theory \cite{Rastall:1973nw}. In the frame of Rastall theory, the covariant divergence of the stress-energy momentum tensor is proportional to the covariant divergence of the curvature scalar, i.e. ${T^\alpha}{}_{\beta;\alpha}\propto R_{; \beta}$. Thus, any solution that has a zero or constant Ricci scalar Rastall theory will be identical to Einstein GR. Explaining the behavior of the new source of Rastall's theory is not an easy task. We can consider, phenomenologically, this new source as an appearance of quantum effects in the classical frame \cite{Fabris:2014xpa}. It is interesting to mention that the topic of non-conservation of  $T^{\alpha \beta}$ is a feature that exists in diffusion models \cite{Calogero:2011re,Nashed:2011fg,Nashed:2016tbj,Calogero:2013zba,Velten:2014uva}. Also, the non-conservation of the energy-momentum tensor and its link to modified gravitational theories has been analyzed in \cite{Koivisto:2005yk, Minazzoli:2013bva}.  The variational principle in the frame of Riemannian geometry is not held due to the non-conservation of $T^{\alpha \beta}$. Nevertheless, some features like Rastall's theory can also be discovered in the frame of Weyl geometry \cite{Almeida:2013dba}. Moreover, external fields in the Lagrangian could give essentially the same behavior as Rastall's theory  (for discussion of the external field see, for example \cite{Chauvineau:2015cha}). An investigation of Rastall gravity, for an anisotropic star with a static spherical symmetry, has been discussed in \cite{Nashed:2022zyi}. { The study of shadow and energy emission rates for a spherically symmetric non-commutative black
hole in Rastall gravity has been carried out \cite{Ovgun:2019jdo}. The
quasinormal modes of black holes in Rastall gravity in the presence of non-linear electrodynamic sources have been studied \cite{Gogoi:2021dkr}. Moreover,  the quasinormal modes of the massless Dirac field for
charged black holes in Rastall gravity have been discussed \cite{Shao:2020gwr}. In the framework of Rastall gravity, a new black hole solution of the Ay\'on-Beato-Garc\'ia type, surrounded by a cloud of strings, is derived  \cite{Gogoi:2021cbp}. A solution of a static spherically symmetric black hole surrounded by a cloud of strings in the frame of Rastall gravity is derived \cite{Cai:2019nlo}. Also two classes of black hole
(BH) solutions, conformally flat and non-singular BHs, are presented in \cite{Moradpour:2019shq}.  A spherically symmetric gravitational collapse of a homogeneous perfect fluid in Rastall
gravity has been done in \cite{Ziaie:2019jfl}.  Oliveira also presented static and spherically symmetric solutions for the Rastall modification of gravity to describe neutron stars \cite{ Oliveira:2015lka}}. 

In the frame of cosmology, Rastall's theory could degenerate into the $\Lambda$ cosmological dark matter, $\Lambda$CDM, at the background and at first order levels which means that a viable model can be constructed in the frame of this theory. However, a few applications in the domain of astrophysics have been done \cite{Batista:2011nu}. { Also, a study of the generalized Chaplygin gas model to fit observations has been carried out in Rastall theory \cite{Fabris:2012hw}. The quantum thermodynamics of the Schwarzschild-like black hole found in the bumblebee gravity model has been discussed in \cite{Kanzi:2019gtu}.}  In recent years, various BH solutions, and in particular, BH solutions of the Rastall field equations, have been investigated in many scientific research papers. Among these are  charged static spherically
symmetric BH solutions \cite{Bronnikov:2016odv, Heydarzade:2016zof}, Gaussian BH solutions \cite{Spallucci:2017mto,Ma:2017jko}, rotating BH solutions \cite{Kumar:2017qws,Xu:2017vse},
Abelian-Higgs strings \cite{deMello:2014hra}, G$\ddot{o}$del-type BH solutions \cite{Santos:2014ewa}, black branes \cite{Sadeghi:2018vrf}, wormholes \cite{Moradpour:2016ubd}, BH  solutions surrounded by fluid, electromagnetic field \cite{Heydarzade:2017wxu} or
quintessence fluid \cite{MoraisGraca:2017hrf}, BH thermodynamics \cite{Lobo:2017dib},
among other theoretical efforts \cite{Licata:2017rfx,Darabi:2017coc,Carames:2015tpa,Salako:2016ihq}.   It is the aim of the present study to show the effect of the Rastall parameter in the domain of spherically symmetric spacetime using a special form of NED coupled with Einstein's GR. 

  This paper has the following structure: in the next section,  we present a summary of Rastall's theory. In Subsection \ref{S31} we give the NED field equations of Rastall's gravity, then we apply them to a spherically symmetric spacetime with two unequal metric potentials and derive the NED differential equations. We solve this system and derive a new BH  solution that involves Rastall's parameter.  In Subsection \ref{S4}, we extract the physical properties of the BH solution and show that the metric potentials asymptote as  Anti-de-Sitter (A)dS Reissner--Nordstr\"om.  Despite we applied the NED field equations without cosmological constant, we get (A)dS Reissner--Nordstr\"om.  This means that the Rastall parameter acts as a  cosmological constant in this special form of NED theory. This result is consistent with the study given by Visser \cite{Visser:2017gpz}.    It is important to stress that this solution in the GR limit, i.e., when the Rastall parameter equals zero,  coincides with the Reissner-Nordstr"om solution.  In Subsection \ref{S5} we derive the stability of geodesic motion using geodesic deviations. In Section \ref{S6}, we study some thermodynamical quantities. In Subsection  \ref{S61}, we show that our BH satisfies the first law of thermodynamics. In Section \ref{S8} we discuss the output results of this study.
%%%%%%%%%%%%%%%%%%%%%%%%%%%%%%%%%%% Section 3 %%%%%%%%%%%%%%%%%%%%%%%%%%%%%%%%%%%%%%%%
\section{spherically symmetric BH solution  }\label{S3}
%%%%%%%%%%%%%%%%%%%%%%%%%%%%%%%%%%%%%%%%%%%%%%%%%%%%%%%%%%%%%%%%%%%%%%%%%%%%%%%%%%%%%%
Rastall's assumptions  \cite{Rastall:1973nw,Rastall:1976uh},
for a spacetime with a Ricci scalar $R$ filled by an energy-momentum  $T_{\mu \nu}$, we have
 \begin{equation}\label{e1}
{T^{\alpha \beta}}_{; \alpha}=\epsilon {\mathcal{R}\,_;}{}\,^\beta\,,
\end{equation}
where $\epsilon$ is the Rastall parameter, which is responsible for the deviation from the standard GR conservation law. Equation (\ref{e1}) returns to Einstein's GR when the Ricci scalar is vanishing or has a constant value.

 Using the above data, we can write  Rastall field equations in the form \cite{Rastall:1973nw,Rastall:1976uh}:
 \begin{equation} \label{e2}
\mathcal{R}_{\alpha \beta}-\left[\frac{1}{2}+\lambda\right] g_{\alpha \beta} \mathcal{R}=\chi T_{\alpha \beta}\,,
 \end{equation}
 where $\lambda=\chi \epsilon$ and $\chi$ is the Newtonian gravitational constant  and the units are used so that the speed of light $c = 1$. Here  ${\cal R}_{\alpha \beta}$ is
the Ricci tensor, ${\cal R}$ is the  Ricci scalar, $g_{\alpha \beta}$ is the metric
tensor, and $T_{\alpha \beta}$ is the energy-momentum tensor describing the material content. The  constant $\epsilon$ is the Rastall parameter that is responsible for the
deviation from GR and  when  ($\epsilon = 0$) we get GR theory.

 {The modification in the spacetime geometry given by the L.H.S. of Eq. (\ref{e2}) links to two  modifications of different   material
content of the right hand side of Eq. (\ref{e2}):\vspace{0.1cm}\\ (i) Firstly, Eq. (\ref{e2}) is mathematically
equivalent to adding  new material
of the actual material sources to the right-hand side
of the standard GR field equations, which  can be
 seen  as an effective source accompanying the actual
material sources considered in the model. For this reason, we can rewrite
Eq. (\ref{e2})  in a mathematical  equivalent form as \cite{Rastall:1973nw,Rastall:1976uh}:
\begin{equation} \label{e3}
\mathcal{R}_{\alpha \beta}-\frac{1}{2}g_{\alpha \beta} \mathcal{R}=\chi T_1{_{\alpha \beta}}\,, \qquad \textrm{where}\qquad  T_1{_{\alpha \beta}}= T{_{\alpha \beta}}-\frac{ \chi\,\epsilon}{1+4\chi\, \epsilon}g_{\alpha \beta}T\,.
 \end{equation}
The term $-\frac{\epsilon}{1+4\epsilon}g_{\alpha \beta}T$ is the energy-momentum tensor that represents the effective source that arises from
the actual material and $T$ is the trace of $T_{\alpha \beta}$, i.e., $T=g_{\alpha \beta}T^{\alpha \beta}=-(1+4\epsilon)R$.\\
Now rewrite Eq. (\ref{e3}) in the form:\footnote{{ In this study we assume  the relativistic units, i.e., $\chi=\frac{8\pi G}{c^4}=1$.}}
\begin{equation} \label{e33}
{ {\mathcal{R}_{\alpha \beta}-g_{\alpha \beta} \mathcal{R}}\left[\frac{1}{2}+ \epsilon\right]\equiv {\mathcal{R}_{\alpha \beta}+g_{\alpha \beta} {T}}\left[\frac{ 1+2\epsilon}{2(1+4\epsilon)}\right] =T_{\alpha \beta}\,.}
 \end{equation}
 In this study,  we will use Eq. (\ref{e33}) but we will assume the energy-momentum tensor $T_{\alpha \beta}$ to be { combined} with electromagnetic field and takes the following form:}
\begin{equation}\label{e4}
{{T}}_{\alpha \beta}=E_{\alpha \beta}, \qquad \textrm{where} \qquad E_{\alpha \beta}={F^\mu}{}_\alpha F_{\mu \beta}-\frac{1}{4}g_{\alpha \beta}F\,,
\end{equation}
with   $F_{\mu \beta}$ being the  antisymmetric Faraday tensor and $F=F^{\mu \nu}F_{\mu \nu}=d\xi$ and $\xi=\xi_\alpha dx^\alpha$ is the electromagnetic  gauge potential Maxwell field \cite{Capozziello:2012zj}. The tensor $F_{\mu \beta}$  satisfies the vacuum Maxwell equations
\begin{equation}\label{e5}
{F^{\alpha \beta}}_{; \alpha}=0\,, \qquad \qquad F_{\alpha \beta;\sigma}+F_{ \beta \sigma;\alpha}+F_{ \sigma \alpha;\beta }=0\,.
\end{equation}

ii){ - Secondly}, this modification implies a violation of the local conservation of the tensor  $ T_1{_{\alpha \beta}}$ of an
actual material source because its divergence is not necessarily to be vanishing.

{ It is important to stress that Eq. (\ref{e33}) with the energy-momentum tensor given by Eq. (\ref{e5}) has a contradiction since the LHS of Eq. (\ref{e33}) has a non-vanishing covariant derivative,   $\left\{{\mathcal{R}^{\alpha \beta}-g^{\alpha \beta} \mathcal{R}}\left[\frac{1}{2}+\frac{ \epsilon}{1+4\epsilon}\right]\right\}_{;\beta}\neq 0$ while the RHS has a vanishing value, ${{{T}}^{\alpha \beta}}_{;\beta}=0$. Thus, the only way to overcome this issue is the fact that the solution of these field equations must have a zero Ricci scalar\footnote{ In the frame of Rastall theory, Reissner-Nordstr\"om is a solution since its Ricci scalar has a vanishing value.}  which ensures the well-known results in the literature that the Rastall parameter has no effect in the linear Maxwell field.}
%%%%%%%%%%%%%%%%%%%%%%%%%%%%%%%%%%% Section 3 %%%%%%%%%%%%%%%%%%%%%%%%%%%%%%%%%%%%%%%%
\subsection{Nonlinear charged spherically symmetric BH solution in Rastall's theory}\label{S31}
%%%%%%%%%%%%%%%%%%%%%%%%%%%%%%%%%%%%%%%%%%%%%%%%%%%%%%%%%%%%%%%%%%%%%%%%%%%%%%%%%%%%%%

In this subsection, we are going to present a special form of NED theory coupled with GR. For this aim, we are going to take into account a dual representation, i.e.,  imposing the auxiliary field ${\cal S}_{\alpha \beta}$, which is convenient to couple with GR  \cite{1999PhLB..464...25A,1987JMP....28.2171S}. Especially, we
involve the   Legendre transformation
\begin{equation} \label{Ha}
{\mathbb{H}}=2{ F} { L}_{ F}-{ L},
\end{equation}
where  $\mathbb{H}$ is an arbitrary function,  ${ L}_{ F}\equiv\frac{\partial { L}}{\partial { F}}$  and  { ${ L({ F})}$ is an arbitrary function of $F$. if  ${ L({ F})}={ F}$ we return to the linear case}.
Assuming
\begin{eqnarray}\label{rel22vvv}
\label{rel22}
{\cal S}_{\mu \nu}={ L}_{ F} { F}_{\mu \nu}\,, \qquad
{\cal S}=\frac{1}{4}{\cal S}_{\alpha \beta}{\cal
S}^{\alpha \beta}={ L}_{F}^2 {F}\,, \qquad  \textrm{with} \qquad { F}_{\mu \nu}={\mathbb{H}}_{ \cal S} {\cal
S}_{\mu \nu}\,,  \end{eqnarray}
where ${\mathbb{H}}_{\cal
S}=\frac{\partial {\mathbb{H}}}{\partial {\cal S}}$.  The field equation of   nonlinear electrodynamics yields the form
\cite{1999PhLB..464...25A}:
\begin{equation} \label{maxf}
\partial_\nu \left( \sqrt{-g} {\cal S}^{\mu \nu} \right)=0,\end{equation}
where the energy-momentum tensor  of the NED is defined as:
\begin{equation} \label{max1}
{{{{
T^\nu{}_\mu}}^{{}^{{}^{^{}{^{}{\!\!\!\!\scriptstyle{\hspace{0.05cm} NED}}}}}}}}\equiv
2({\mathbb{H}}_{ S}{S}_{\mu \alpha}{ S}^{\nu \alpha}-\delta_\mu^\nu
[2{ S}{\mathbb{H}}_{ S}-{\mathbb{H}}]).
\end{equation}
{ We mention that in   general   Eq.  (\ref{max1}) has  a non-vanishing  trace\footnote{The non-vanishing of the trace is an important property in the frame of Rastall's theory so that the effect of Rastall parameter maybe appear unlike Maxwell field theory.}
  \begin{equation}
 {{{{
T}}^{{}^{{}^{^{}{^{}{\!\!\!\!\scriptstyle{\hspace{0.15cm} NED}}}}}}}} =8({\mathbb{H}}-{\mathbb{H}}_{\cal
S}{\cal  S})\neq 0\,,
\label{traceEM}
\end{equation}
and  has a vanishing value in the linear theory, i.e., { when $\mathbb{H}=F$} and ${ \cal S}=F$.}
  Finally, the electric and magnetic fields in the NED case take the form  \cite{1999PhLB..464...25A,1987JMP....28.2171S}:
\begin{eqnarray} \label{Max3}
&&E = \int{F}_{tr}dr =\int{\mathbb{H}}_{S}{S}_{tr}dr\,,
\qquad \qquad  \qquad B_r =\int {F}_{r\phi}d\phi =\int{\mathbb{H}}_{S}{S}_{r\phi}d\phi,
\nonumber\\
&&
B_\theta = \int{ F}_{\theta r}dr
=\int{\mathbb{H}}_{S}{S}_{\theta r}dr, \qquad \qquad \qquad  B_\phi=\int{ F}_{\phi r}dr =\int{\mathbb{H}}_{S}{S}_{\phi r}dr\,,
\end{eqnarray}
where $E$ and B are the components of the electric and magnetic fields respectively.
Now we are going to use the field equation (\ref{e33}) with the energy-momentum tensor $T_{\alpha \beta}$, that is combined with the NED, and get:
\begin{equation}\label{e6}
{{{
T_{\mu \nu}}^{{}^{{}^{^{}{^{}{\!\!\!\!\scriptstyle{\hspace{0.05cm} NED}}}}}}}}\equiv{\mathbb E}_{\mu \nu}, \qquad \textrm{where} \qquad {{\mathbb E}_\mu{}^ \nu=2({\mathbb{H}}_{ S}{S}_{\mu \alpha}{ S}^{\nu \alpha}-\delta_\mu^\nu
[2{ S}{\mathbb{H}}_{ S}-{\mathbb{H}}])}\,.
\end{equation}

  Now, let us assume that the spherically symmetric spacetime has the form:
 \begin{eqnarray} \label{met12}
& &  ds^2=-\mu(r)dt^2+\frac{dr^2}{\nu(r)}+r^2(d\theta^2+{ \sin^2\theta}\,d\phi^2)\,,  \end{eqnarray}
where $\mu(r)$ and $\nu(r)$  are  unknown  functions of the radial coordinate { $r$}. For the spacetime (\ref{met12}), the symmetric affine connection takes the form:
 \begin{eqnarray} \label{cons}
&&{\Gamma_{tt}}^r=\frac{1}{2}\nu \mu'\,, \qquad \qquad {\Gamma_{tr}}^t=\frac{\mu'}{2\mu} \,, \qquad \qquad {\Gamma_{rr}}^r=\frac{\nu'}{2\nu} \,, \qquad \qquad  {\Gamma_{r\theta}}^\theta={\Gamma_{r\phi}}^\phi=\frac{1}{r} \,,\nonumber\\
&& {\Gamma_{\theta \theta}}^r=-r\nu \,, \qquad \qquad {\Gamma_{\theta \phi}}^\phi=\cot\theta \,, \qquad \qquad {\Gamma_{\phi \phi}}^r=-r\nu \sin^2\theta\,, \qquad  \qquad {\Gamma_{\phi \phi}}^\theta=-\sin\theta\, \cos\theta\,.
 \end{eqnarray}
The Ricci scalar of the spacetime (\ref{met12}) has the form:
  \begin{eqnarray} \label{Ricci}
  \mathrm{ R(r)}=\frac{r^2\nu \mu'^2-r^2 \mu \mu' \nu'-2r^2 \mu \nu \mu''-4r\mu [\nu \mu'-\mu \nu']+4\mu^2(1-\nu)}{2r^2\mu^2}\,.
  \end{eqnarray}
  Here, $\mu\equiv \mu(r)$,  $\nu\equiv \nu(r)$,  $\mu'\equiv \frac{d\mu}{dr}$, $\mu''\equiv \frac{d^2\mu}{dr^2}$ and $\nu'\equiv \frac{d\nu}{dr}$.
 \\

Using Eq. (\ref{met12}) in Eq.  (\ref{e33}), where the energy-momentum tensor is given by Eq. (\ref{e6}),
% and using the nonlinear Maxwell equation given by Eq. (\ref{maxf})
we get:
\begin{eqnarray}
&& \textrm{ The { $t\,t$- component} of Rastall field equation is:}\nonumber\\
&&\frac{1}{2r^2\mu^2[2\mu \nu \xi''+\xi'(\mu\nu'-\nu\mu')]}\Bigg\{2\mu\nu \xi''\left\{2\mu\nu\epsilon r^2\mu''-r^2\nu\epsilon \mu'^2+r\mu\mu' \epsilon [4\nu+r\nu']+2\mu^2[(1+2\epsilon)[r\nu'+\nu]+1+r^2\mathbb {H}+2\epsilon]\right\}\nonumber\\
&&+\xi'\Bigl(2r^2\mu \nu \mu''\epsilon[\mu\nu'-\nu\mu']+r^2\nu^2\epsilon \mu'^3-2r\mu \nu \mu'^2[r\nu'+2\nu]\epsilon+\mu^2\mu'\left[r^2\epsilon \nu'^2-2r\nu\nu'-2\nu(\{1+2\epsilon\}(\nu-1)-r^2\mathbb {H})\right]\nonumber\\
&&+\mu^3\left\{2\nu'[(1+2\epsilon)\{r\nu'+\nu-1\}-r^2\mathbb{H}]+r^2\nu \mathbb{H}'\right\}\Bigr)\Bigg\}=0\,,\nonumber\\
&& \textrm{ The { $r\,r$- component}  of  Rastall field equation is:}\nonumber\\
&&\frac{1}{2r^2\mu^2[2\mu \nu \xi''+\xi'(\mu\nu'-\nu\mu')]}\Bigg\{2\mu\nu \xi''\left\{2\mu\nu\epsilon r^2\mu''-r^2\nu\epsilon \mu'^2+r\mu\mu'  [2(1+2\epsilon)\nu+\epsilon r\nu']+2\mu^2[(1+2\epsilon)[\nu-1]+2\epsilon r\nu'-r^2\mathbb {H}]\right\}\nonumber\\
&&+\xi'\Bigl(2r^2\mu \nu \mu''\epsilon[\mu\nu'-\nu\mu']+r^2\nu^2\epsilon \mu'^3-2r\mu \nu \mu'^2[r\nu'\epsilon+(1+2\epsilon)\nu]+\mu^2\mu'\left[r^2\epsilon \nu'^2+2r\nu\nu'-2\nu(\{1+2\epsilon\}(\nu-1)-r^2\mathbb {H})\right]\nonumber\\
&&+2\mu^3\left\{2\epsilon r\nu'^2+\nu'[(1+2\epsilon)\{\nu-1\}-r^2\mathbb{H}]+r^2\nu \mathbb{H}'\right\}\Bigr)\Bigg\}=0\,,\nonumber\\
&& \textrm{ The}\, \mathrm{\theta\, \theta = \phi \, \phi}\textrm{- component of  Rastall field equation is:}\nonumber\\
&&\frac{1}{4r^2\mu^2[2\mu \nu \xi''+\xi'(\mu\nu'-\nu\mu')]}\Bigg\{2\mu\nu \xi''\Bigl\{2\mu\nu(1+2\epsilon) r^2\mu''-r^2\nu(1+2\epsilon) \mu'^2+r\mu\mu'  [2(1+4\epsilon)\nu+(1+2\epsilon) r\nu']+2\mu^2[4\epsilon[\nu-1]\nonumber\\
&&+(1+4\epsilon) r\nu'-2r^2\mathbb {H}]\Bigr\}+\xi'\Bigl(2r^2\mu \nu \mu''(1+2\epsilon)[\mu\nu'-\nu\mu']+r^2\nu^2(1+2\epsilon)\mu'^3-2r\mu \nu \mu'^2[r\nu'(1+2\epsilon)+(1+4\epsilon)\nu]\nonumber\\
&&+\mu^2\mu'\left[r^2(1+2\epsilon) \nu'^2-4\nu(2\epsilon(\nu-1)-r^2\mathbb {H})\right]+2\mu^3\left\{(1+4\epsilon) r\nu'^2+2\nu'[2\epsilon\{\nu-1\}-r^2\mathbb{H}]+4r^2\nu \mathbb{H}'\right\}\Bigr)\Bigg\}=0\,,\nonumber\\
\label{feqcn}
\end{eqnarray}
where $\mathbb{H}$ is an arbitrary  function and $\xi$ is the field of electric charge.
{ Equations (\ref{feqcn}) reduce to the linear charged  Einstein's  field equations} when $\epsilon=0$ and $\mathbb {H}=F$ \cite{Guo:2021bwr,Nashed:2020kdb}.
The exact solution of  Eq. (\ref{feqcn}) for the electric field  takes the form:\footnote{Solution (\ref{sol1cn}) has been checked using Maple software 19.}:
\begin{eqnarray}\label{sol1cn}
&&{ \mu(r)=\frac{c_2(c_3r^4+(1+4\epsilon)[12r^2+12rc_4-3c_5])}{r^2}}\,, \qquad \qquad { \nu(r)=\frac{c_3r^4+(1+4\epsilon)[12r^2+12rc_4-3c_5]}{12r^2(1+4\epsilon)}}\,,\nonumber\\
&&{  \xi(r)=\frac{c_1}{r}\,, \qquad \qquad {\mathbb{H}=c_3+\frac{c_5}{r^4}\equiv c_3+F\,.} }
\end{eqnarray}
%\begin{equation}\label{sol1}
%\mu(r)=\frac{r^2c_2}{12}+(1+4\epsilon)[1+\frac{c_3}{r}-\frac{c_4}{4r^2}]\,, \nu=\qquad \qquad \xi(r)=\frac{c_1}{r^2}\,, \qquad \qquad \mathbb{H}=c_2+\frac{c_4}{r^4}\,.
%\end{equation}
{  The Rastall parameter has an effect in the NED case, as shown by Eq.~\eqref{sol1cn}. We return to the linear charged case  when $\mathbb{H}=F\equiv\frac{c_5}{r^4}$  \cite{Prihadi:2019isb}. We stress the fact that if we repeat the same above calculations taking into account the electric and magnetic fields, given by Eq. (\ref{Max3}), we can easily verify the same conclusion of the above case, i.e.,  Rastall's parameter has an effect and its behavior will be similar to the form given by Eq. (\ref{sol1cn}). If we like to derive a solution that is different from Einstein's GR, we must generalize Rastall's theory to $f(R)$-Rastall's theory \cite{Shahidi:2021lxt}}
 %In conclusion, we have studied the NED  case and get a BH solution that has a Rastall parameter.
%%%%%%%%%%%%%%%%%%%%%%%%%%%%%%%%%%% Section 3 %%%%%%%%%%%%%%%%%%%%%%%%%%%%%%%%%%%%%%%%
\subsection{The physical properties  of the BH solutions (\ref{sol1cn})}\label{S4}
%%%%%%%%%%%%%%%%%%%%%%%%%%%%%%%%%%%%%%%%%%%%%%%%%%%%%%%%%%%%%%%%%%%%%%%%%%%%%%%%%%%%%%
Now we are going  to explain the physics of the BH solution (\ref{sol1cn}). For such purposes, we rewrite the components of the metric potential of the  BH   (\ref{sol1cn}) as:
\begin{eqnarray} \label{ass2n}
&& \mu(r)=\nu(r)=r^2\Lambda_{eff}+1-\frac{2M}{r}+\frac{q^2}{r^2}\,, \qquad \xi=-\frac{q}{r}\,, \qquad \mathbb{H}=12\Lambda_{eff.}(1+4\epsilon)-\frac{4q^2}{r^4}\,.
\end{eqnarray}
where we have put
\begin{eqnarray} \label{ass2n1} c_1=-q\,, \quad c_2=\frac{1}{12(1+4\epsilon)}\,, \quad \Lambda_{eff.}=c_3c_2\,, \quad c_4=-2M\,, \qquad  \textrm{and} \quad  -4q^2=c_5\,, \quad c_5=-4\sqrt{c_1}\,.\end{eqnarray}
{ Equation (\ref{ass2n1})} shows that we have got an effective cosmological constant in the solution of the NED charged case while their field equations have no cosmological constant. This means that the Rastall parameter acts as an effective cosmological constant in the NED charged case with the fact that the Rastall parameter  $\epsilon \neq -\frac{1}{4}$. From Eq.~(\ref{ass2n}) and Eq.~(\ref{met12}) we
get \footnote{{This result is consistent with what  have done in \cite{Visser:2017gpz} where the author has shown that the Rastall theory is equivalent to Einstein's general relativity or equivalent to Einstein's field equation plus an arbitrary cosmological constant}}:
\begin{align}
\label{metafn}
ds^2=& -\left\{r^2\Lambda_{eff}+1-\frac{2M}{r}+\frac{q^2}{r^2}\right\}dt^2+\frac{dr^2}{r^2\Lambda_{eff}+1-\frac{2M}{r}+\frac{q^2}{r^2}}+d\Omega^2
\,,
\end{align}
where $d\Omega^2= r^2(d\theta^2+{ \sin^2\theta}\,d\phi^2)$ is a 2-dimensional unit sphere.

{ Eq.~(\ref{metafn}) shows that solution (\ref {sol1cn}) asymptotes as (A)dS and does
not equal to  Reissner--Nordstr\"om  spacetime due to Rastall parameter. Equation (\ref{metafn})
investigates clearly that the Rastall parameter acts as a cosmological constant. Equation  (\ref{metafn}) coincides with GR when $\mathbb{H}=F$ which means $c_3=0$  and this gives Rissner-Nordstr\"om BH solution because $\Lambda_{eff}=0$ \cite{Bravo-Gaete:2022mnr,Alvarez:2022upr1}. }
From  Eq.~(\ref{ass2n}) and Eq.~(\ref{Ricci}) we get:
\begin{align}
\label{R1}
\mathrm{R}(r)=-12\Lambda_\mathrm{eff}\,.
\end{align}
 Equation~(\ref{R1}), shows in a clear way that the Rastall parameter acts as a cosmological constant and the conservation law of both sides of Eq. (\ref{e2}) are satisfied.

Using Eq.~(\ref{ass2n}) we get the invariants  as:
\begin{align} \label{invn}
&&\mathcal{R}_{\mu \nu \rho \sigma} \mathcal{R}^{\mu \nu \rho \sigma}=-24\Lambda_\mathrm{eff}+\frac{48m^2}{r^6}-\frac{96mq^2}{r^7}+\frac{56q^4}{r^8}\,,\qquad \mathcal{R}_{\mu \nu } \mathcal{R}^{\mu \nu }=36\Lambda_\mathrm{eff}+\frac{4q^4}{r^8}\,,\qquad  \mathcal{R}= -12\Lambda_\mathrm{eff}\,.
\end{align}
Here $\left( \mathcal{R}_{\mu \nu \rho \sigma} \mathcal{R}^{\mu \nu \rho \sigma},
\mathcal{R}_{\mu \nu} \mathcal{R}^{\mu \nu}, \mathcal{R} \right)$  are
the Kretschmann scalar, the Ricci tensor square,{ and} the Ricci scalar, respectively. The Kretschmann scalar and the Ricci tensor square have
a true singularity when $r=0$. All the above invariants are identical with the invariant of  { $(A)dS$}-Reissner--Nordstr\"om   BH solution of GR. The discussion of the invariant of  (A)dS Reissner--Nordstr\"om can be applied on the invariant given by Eq. (\ref{invn}) with the exclusion of the value $\epsilon=-\frac{1}{4}$.

{Before we close this subsection we are going to calculate the trace of the NED given by Eq.~(\ref{traceEM}) using solution (\ref{sol1cn}) and get:
\begin{eqnarray}\label{trac}
 {{{{
T}}^{{}^{{}^{^{}{^{}{\!\!\!\!\scriptstyle{\hspace{0.15cm} NED}}}}}}}}=c_3\neq 0\,.
\end{eqnarray}
Equation (\ref{trac}) shows in a clear way that if $c_3=0$ we will get a vanishing trace and in that case, Rastall's parameter will have no effect which supports the above discussion.}

%%%%%%%%%%%%%%%%%%%%%%%%%%%% Section 7 %%%%%%%%%%%%%%%%%%%%%%%%%%%%%

 \subsection{Stability of geodesic motion { of BH given by Eq.~(\ref{ass2n})}}\label{S5}

%%%%%%%%%%%%%%%%%%%%%%%%%%%%%%%%%%%%%%%%%%%%%%%%%%%%%%%%%%%%%%%%%%%%

\begin{figure}[ht]
\centering
\subfigure[~Stability of the BH for the case $\Lambda_{eff}= 0$ and $\Lambda_{eff.}=0.083$]{\label{fig:1}
\includegraphics[scale=0.25]{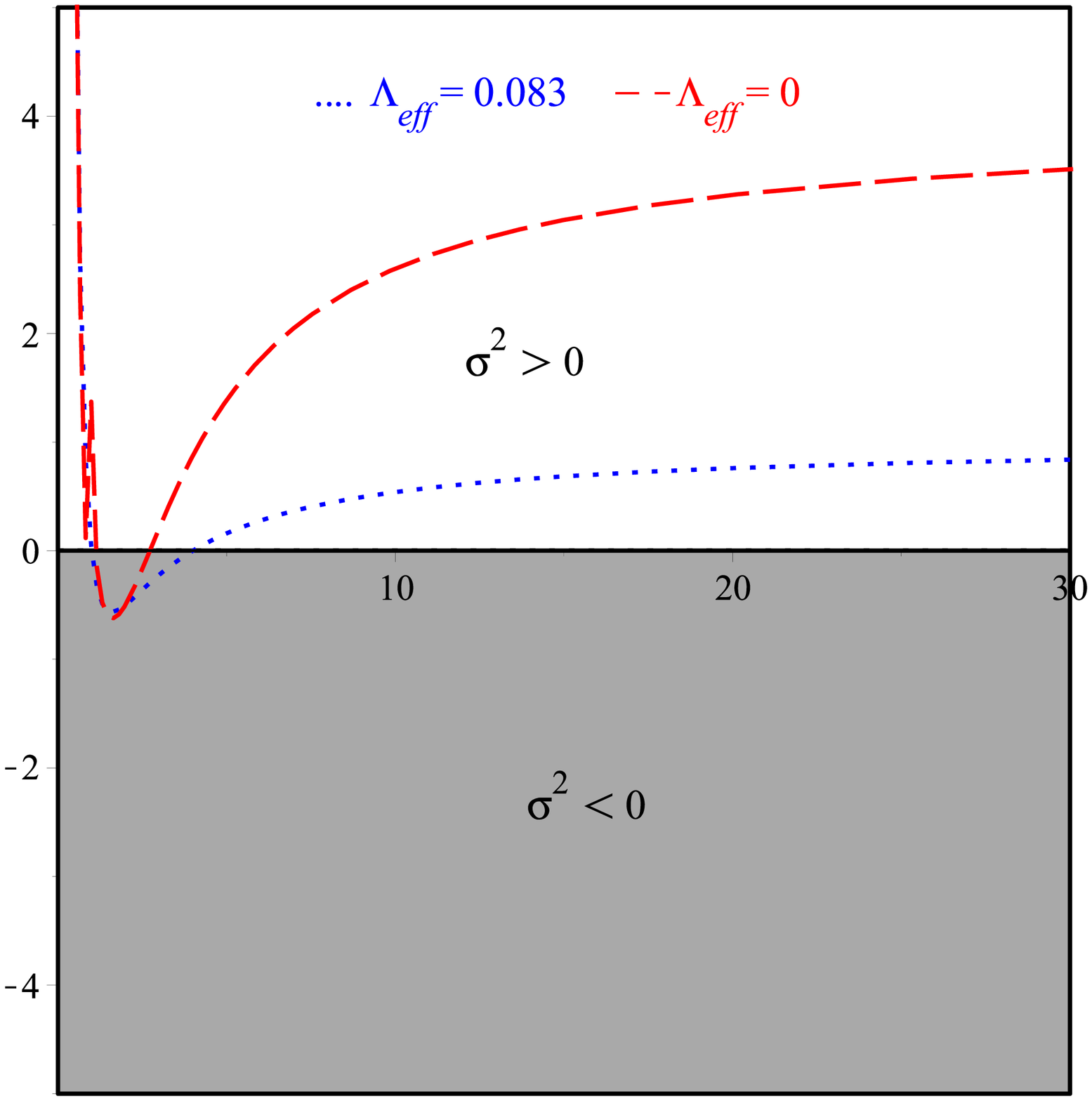}}\hspace{0.2cm}
\subfigure[~Horizons of the  linear Maxwell field ]{\label{fig:1a}\includegraphics[scale=0.25]{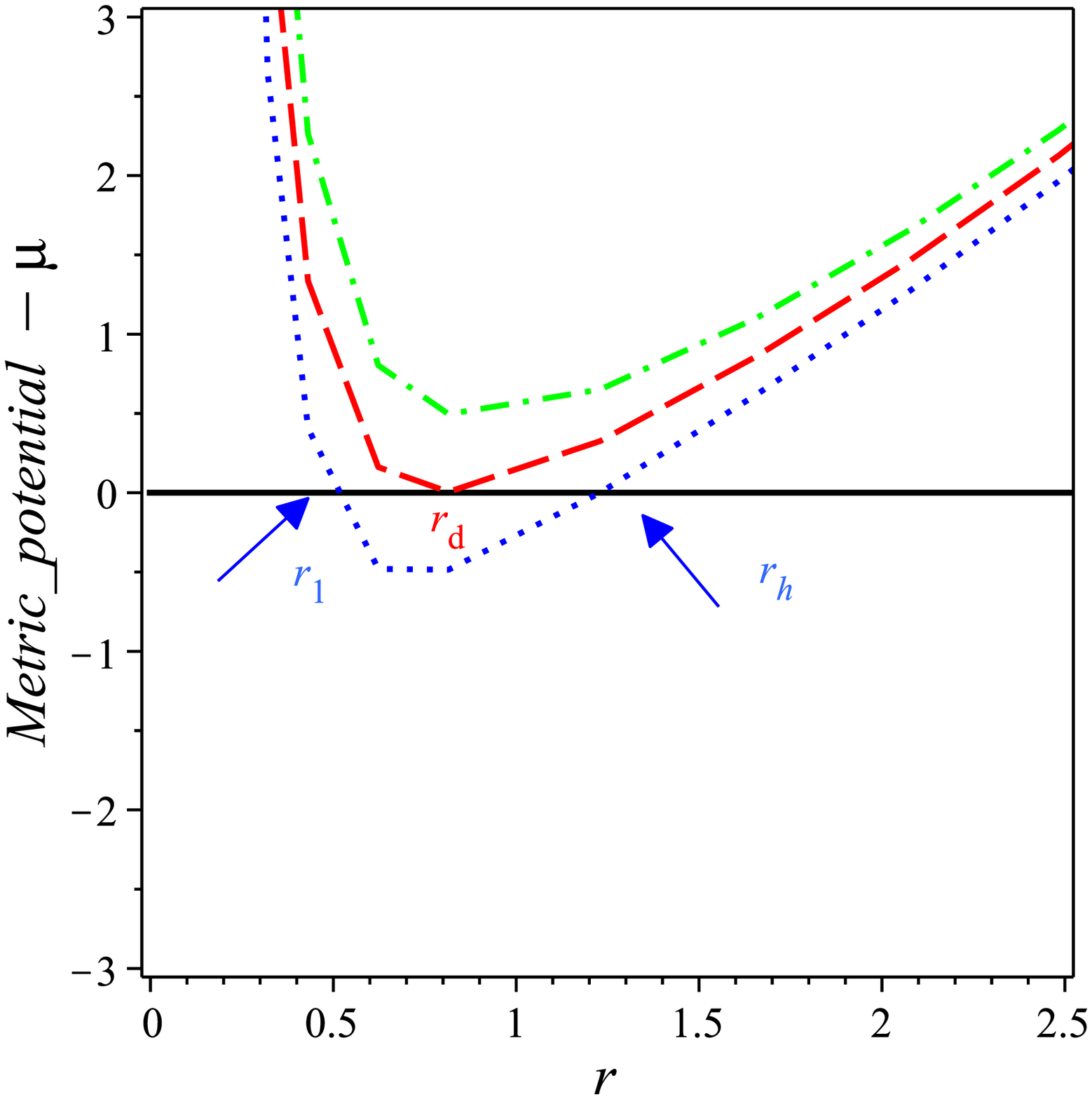}}\hspace{0.5cm}
\subfigure[~Horizons of the non--linear Maxwell field ]{\label{fig:1b}\includegraphics[scale=0.25]{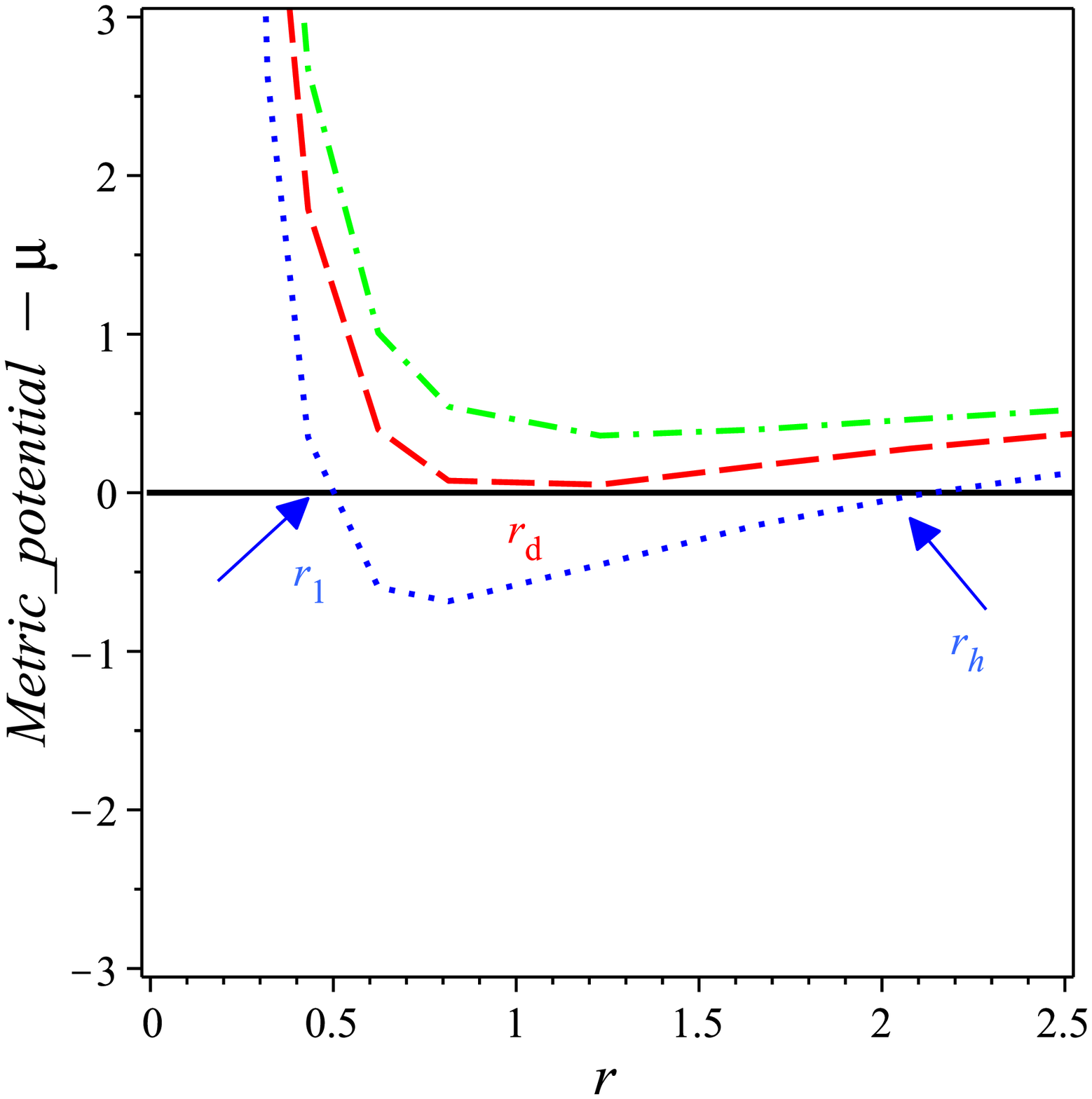}}
\caption{Plot \subref{fig:1} shows the behavior of Eq.~(\ref{stab1}) viz  $r$ for BH (\ref{ass2n}). The behavior of  the metric potential $\mu(r)$ which characterizes the  horizons by putting $\mu(r)=0$: \subref{fig:1a} for linear Maxwell Rastall gravity theory;  \subref{fig:1b} for the nonlinear electrodynamics Rastall's theory. The values of $m$ for the linear case are 1.3; 0.99; 0.8 and q=1 while for the nonlinear case m=1.3; 1.1 and 0.9,  q=1 and $\Lambda_{eff.}$=0.3}
\label{Fig:1}
%\caption{ {}}
%\label{Fig:1}
\end{figure}

{  The equations of geodesic are given by \cite{Misner:1974qy}:
\begin{equation}
\label{ge3}
\frac{d^2 x^\gamma}{d\varepsilon^2}
+ \left\{ \begin{array}{c} \gamma \\ \beta \rho \end{array} \right\}
\frac{d x^\beta}{d\varepsilon} \frac{d x^\rho}{d\varepsilon}=0\, ,
\end{equation}
where $\varepsilon$ is  { a canonical
parameter}. Moreover, the equations of
geodesic deviation  are given as \cite{1992ier..book.....D,Nashed:2003ee}:
\begin{equation}
\label{ged333}
\frac{d^2 {\varrho}^\sigma}{d\varepsilon^2}
+ 2\left\{ \begin{array}{c} \sigma \\ \mu \nu \end{array} \right\}
\frac{d x^\mu}{d\varepsilon} \frac{d {\varrho}^\nu}{d\varepsilon}
+ \left\{ \begin{array}{c} \sigma \\ \mu \nu \end{array} \right\}_{,\, \rho}
\frac{d x^\mu}{d\varepsilon} \frac{d x^\nu}{d\varepsilon}{\varrho}^\rho=0\, ,
\end{equation}
where ${\varrho}^\alpha$ is  the deviation of the 4-vector.
Following the procedure done in \cite{Nashed:2021pah,Nashed:2021hgn} one can get the stability condition as:
\begin{equation}
\label{con111}
\frac{3\mu \nu \mu'-\sigma^2 \mu \mu'-2r\nu\mu'^{2}+r\mu\nu \mu'' }{\mu\nu'}>0\, ,
\end{equation}
where $\mu$ and $\nu$ are given by Eq. (\ref{ass2n}).
Using Eq. (\ref{con111}) one can get the following form of $\sigma^2$ as:
\begin{equation}
\label{stab1}
\sigma^2= \frac{3\mu \nu \mu''-2r\nu\mu'^{2}+r\mu \nu \mu'' }{\mu^2 \nu'^2}>0\, .
\end{equation}
Equation (\ref{stab1}) is plotted in Fig.~\ref{Fig:1} using specific  values of the model. In this figure we study $\Lambda_{eff}= 0$, Reissner-Nordstr\"om GR spacetime  and $\Lambda_{eff}\neq 0$ of the BH solution (\ref{ass2n}).  The two cases display the regions where  BH solution is stable/unstable by unshaded and shaded regions, respectively.}

%%%%%%%%%%%%%%%%%%%%%%%%%%%%%%%%%%% Section 4 %%%%%%%%%%%%%%%%%%%%%%%%%%%%%%%%%%%%%%%%

{\section{The Thermodynamical properties of  the   of BH given by Eq.~(\ref{ass2n})}}\label{S6}

%%%%%%%%%%%%%%%%%%%%%%%%%%%%%%%%%%%%%%%%%%%%%%%%%%%%%%%%%%%%%%%%%%%%%%%%%%%%%%%%%%%%%%

The thermodynamics of BH is considered an interesting topic in physics because it enables us to understand the physics of the solution. Two main approaches have been proposed to understand the thermodynamical quantities of the BHs: The first approach, delivered by Gibbons and Hawking \cite{Hunter:1998qe,Hawking:1998ct} constructed to understand the thermal properties of the Schwarzschild BH through the use of Euclidean continuation. In the second approach, one has to define the gravitational surface from which we can define the Hawking temperature. Then one can be able to study the stability of the BH \cite{Bekenstein:1972tm,Nashed:2010ocg,Bekenstein:1973ur,Gibbons:1977mu,Alvarez:2022upr}.  Here, we are going to follow the second approach to investigate the thermodynamics of the (A)dS BH  obtained in Eq. (\ref{ass2n}) and then analyzed its stability. The physical quantities characterized by the  BH (\ref{ass2n}) are the mass, $m$, the charge, and the effective cosmological constant $\Lambda_{eff.}$.

The horizons of   Eq. (\ref{ass2n}), are calculated by deriving the roots of  $\mu(r) = 0$ which we plot in Figs. \ref{Fig:1}\subref{fig:1a} and \ref{Fig:1}\subref{fig:1b} using specific  values. Plots of Figs. \ref{Fig:1}\subref{fig:1a} and \ref{Fig:1}\subref{fig:1b}  indicate the roots of $\mu(r)$ that fix the horizons of BH~(\ref{ass2n}), i.e.,  $r_1$  and   $r_h$ . We should emphasize that in the linear case, for $m>0$, $q>0$ and $\Lambda_{eff.}=0$, we can show that the two roots can be formed when $m> m_{min}>q$. However, when $m = m_{min}$, we fix the degenerate horizons, i.e., $r_{dg}$, at which $r_1=r_h$,  which is the Nariai BH whose thermodynamics is studied \cite{Myung:2007qt,Kim:2008hm,Myung:2007av}. However, when $m<m_{min}<q$, there is no BH formed which means that we have a naked singularity as shown in Fig. \ref{Fig:1}\subref{fig:1a}. The same discussion can be used for the NED case, where the degenerate horizon is shown in Fig. \ref{Fig:1}\subref{fig:1b} \cite{Di96,Dymnikova:2001fb,Nashed:2004pn, Dymnikova:2018uyo,Shirafuji:1996im,Hayward:2005gi,Kim:2008hm,Myung:2007av, Nicolini:2005vd,Sharif:2011ja}. In this study, we use positive values of the effective cosmological constant because this gives two horizons. Nevertheless, it is important to mention that negative values of the effective cosmological constant create the same pattern, which is characterized by two horizons \cite{Bronnikov:2003yi,Bronnikov:2012mf}.
%\begin{figure*}
%\centering
%\subfigure[~Horizons of the  linear Maxwell field ]{\label{fig:2a}\includegraphics[scale=0.35]{JBMDTmetgr}}\hspace{0.5cm}
%\subfigure[~Horizons of the non--linear Maxwell field ]{\label{fig:2b}\includegraphics[scale=0.35]{JBMDTmet}}
%\caption{Plots of  $\mu(r)$ which characterizes the  horizons by putting $\mu(r)=0$: \subref{fig:2a} for linear Maxwell Rastall gravity theory;  \subref{fig:2b} for the nonlinear electrodynamics Rastall's theory. The values of $m$ for the linear case are 1.3; 0.99; 0.8 and q=1 while for the nonlinear case m=1.3; 1.1 and 0.9,  q=1 and $\Lambda_{eff.}$=0.3}
%\label{Fig:2}
%\end{figure*}
The stability of the BH depends on the sign of the heat capacity $H_c$. Now, we are going to discuss the thermal stability of the BHs through  their behavior of heat capacities \cite{Nouicer:2007pu,DK11,Nashed:2018efg,Chamblin:1999tk}:
\begin{equation}\label{m55}
H_c=\frac{dE_h}{dT_h}= \frac{\partial m}{\partial r_h} \left(\frac{\partial T}{\partial r_h}\right)^{-1}\,,
\end{equation}
where $E_h$ is the energy. If $H_{c} > 0$ or ($H_c < 0$), the BH will  thermodynamically stable or unstable, respectively. To understand this process, we suppose that at some point the BH absorbs more radiation than it emits, which yields positive heat capacity, which means that the mass is indefinitely increased. In contrast, when the BH emits more radiation than it absorbs, this yields a negative heat capacity, which means the BH mass is indefinitely decreasing until it disappears. Therefore, BH  that has negative heat capacity is unstable thermally.

\begin{figure*}
\centering
\subfigure[~ Mass--radius relation]{\label{fig:3a}\includegraphics[scale=0.27]{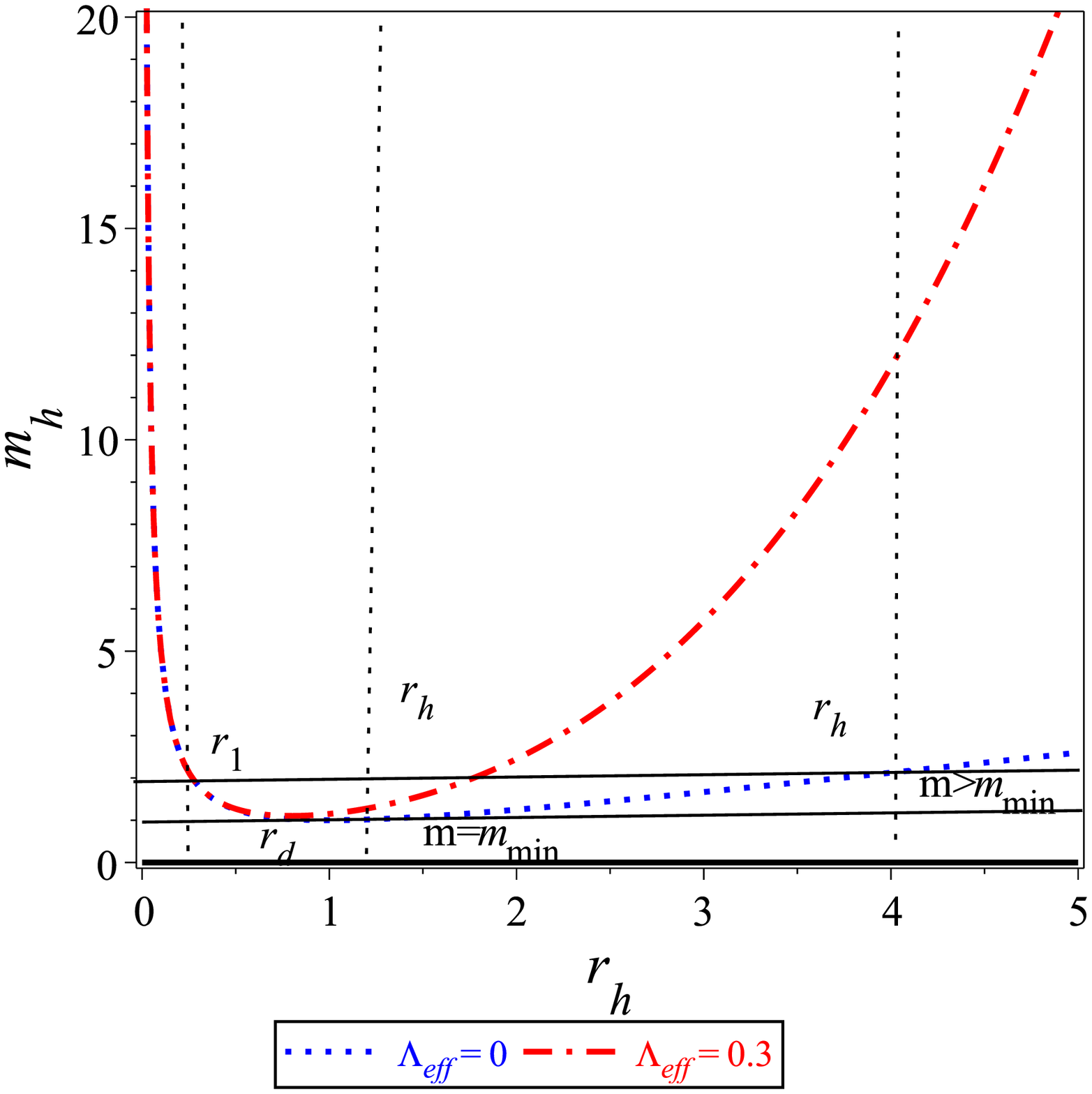}}\hspace{0.2cm}
\subfigure[~Hawking temperature--radius relation]{\label{fig:3b}\includegraphics[scale=0.27]{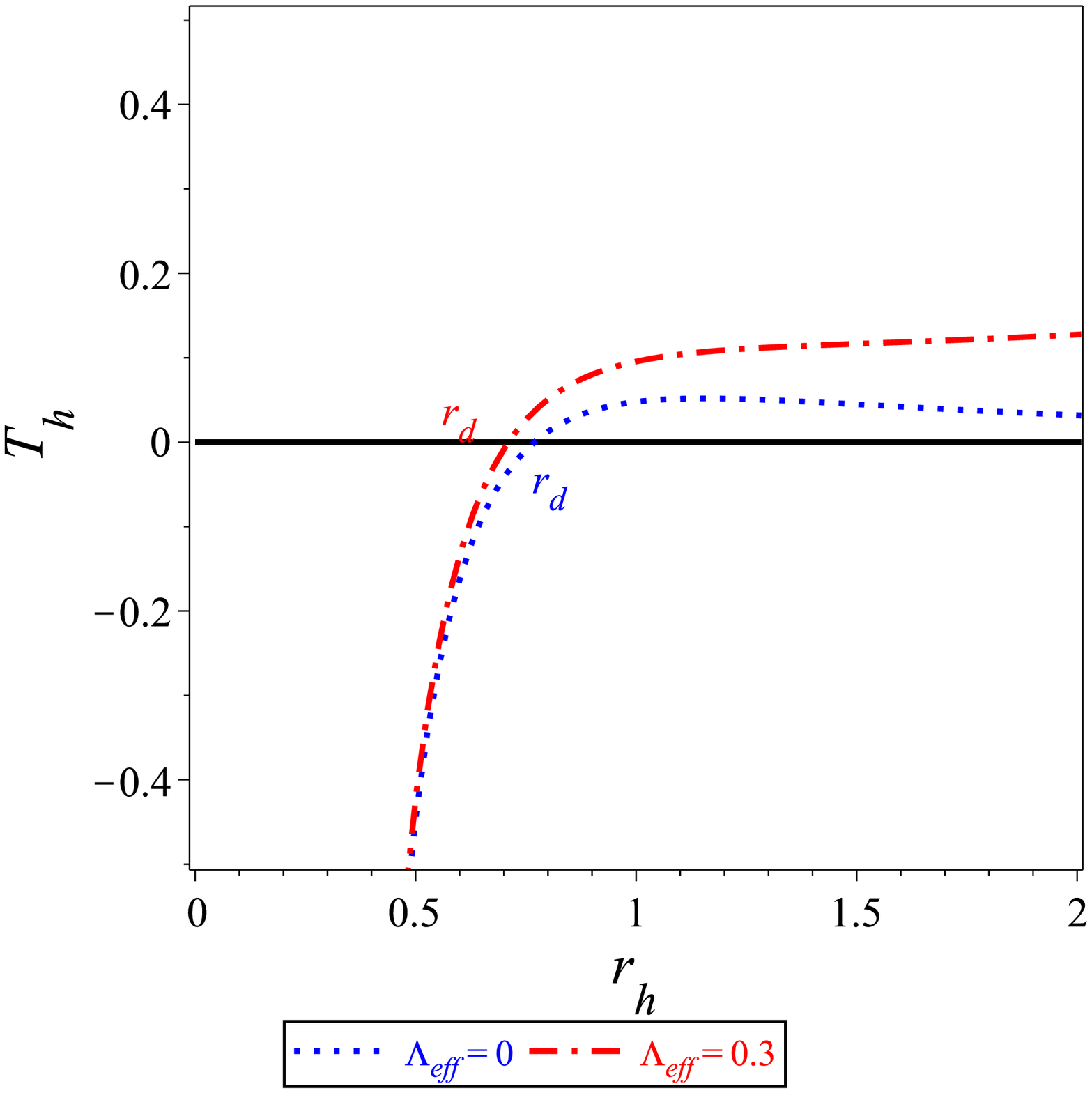}}\hspace{0.2cm}
\subfigure[~Heat capacity--radius relation]{\label{fig:3c}\includegraphics[scale=0.27]{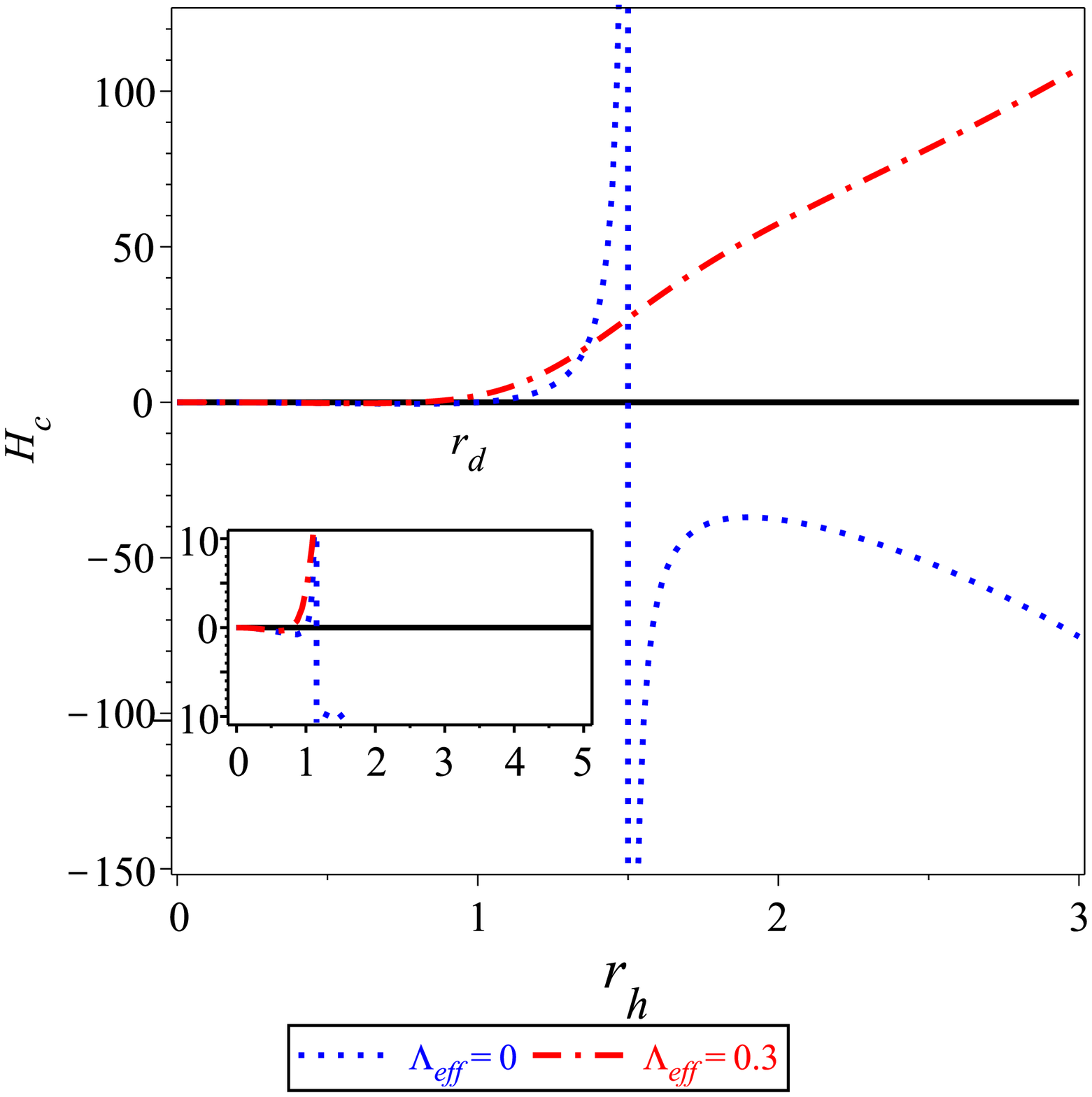}}
\caption{Plots of  thermodynamical quantities of BHs: \subref{fig:3a} The mass-radius relation which determines the minimal mass; \subref{fig:3b} The hawking temperature which  vanishes at $r_h$; \subref{fig:3c} The heat capacity. Moreover, the  linear case investigates a second-order phase transition. All the  figures are plotted  for $m_h=q=1$.}
\label{Fig:3}
\end{figure*}

To calculate Eq. (\ref{m55}), we need the analytical forms of $m_h\equiv m(r_h)$ and $T_h \equiv T(r_h)$. Therefore, let us calculate the mass of the BH  in an event horizon $r_h$. { Thus, we put $\mu(r_h) = 0$, given by Eq.~\eqref{ass2n} and get}:
\begin{eqnarray} \label{m33}
%&&{m_h}_{{}_{{}_{{}_{{}_{\tiny Eq. (\ref{sol1c})}}}}}=\frac{r_h{}^2+q^2}{2r_h}, \nonumber\\
%
&&{m_h}_{{}_{{}_{{}_{{}_{\tiny Eq. (\ref{ass2n})}}}}}=\frac{\Lambda_{eff}r_h{}^4+r_h{}^2+q^2}{2r_h}.
\end{eqnarray}
Equation  (\ref{m33}) shows that the total mass of BH  is  function of $r_h$, the charge and $\Lambda_{eff.}$.  For specific value of the  charge  we plot the relation of the horizon mass-radius in Fig. \ref{Fig:3}\subref{fig:3a} which shows:
\begin{equation} \label{m333}
m(r_h\rightarrow 0)\rightarrow \infty, \qquad \qquad m(r_h\rightarrow \infty)\rightarrow \infty.\end{equation}
%As Fig. \ref{Fig:3}\subref{fig:3a}  shows that for large masses, the double horizons are separated, while for  smaller ones  we do not have horizons. This ensures the results presented in Fig. \ref{Fig:2}\subref{fig:2a}.

%We plot the mass of the black holes within the horizon radius $r_h$ in Fig. \ref{Fig:3}\subref{fig:3a}.
%Notably, the size of the black hole varies between the black hole event horizon $r_{h}$ and the cosmological horizon $r_c$. It has been shown that the radiation from the cosmological horizon $r_c$ is negligible and the black hole evaporates \cite{Fernando:2016ksb,Katsuragawa:2014hda}, when its size approaches the event horizon. However, in the Nariai black hole case, when the two horizons coincide with the degenerate horizon, the temperatures are the same, therefore anti-evaporate may occurs.

The temperature of   BH    is calculated  at the outer event horizon $r = r_h$ as \cite{Hawking:1974sw}:
\begin{equation}
T = \frac{\kappa}{2\pi}\,.
\end{equation}
Here $\kappa$ is the surface gravity defined as $\kappa= \frac{\mu'(r_h)}{2}$. The  temperatures of the BH (\ref{sol1cn}) is given by:
\begin{eqnarray} \label{m44}
%{T_h}_{{}_{{}_{{}_{{}_{\tiny Eq. (\ref{sol1c})}}}}}&=&\frac{1}{2\pi r_h{}^2}\left[m-\frac{q^2}{r_h}\right], \nonumber\\
%
{ {T_h}_{{}_{{}_{{}_{{}_{\tiny Eq. (\ref{ass2n})}}}}}}&=&\frac{1}{2\pi}\left(\Lambda_{eff} r_h+\frac{1}{r_h{}^2}\left[m-\frac{q^2}{r_h}\right]\right)\,,
\end{eqnarray}
with ${T_h}$ being the temperature at $r_h$. For our two cases,  linear and nonlinear electrodynamics,  we depict the temperatures in Fig. \ref{Fig:3}\subref{fig:3b} for specific values. Figure \ref{Fig:3}\subref{fig:3b}  shows that the horizon temperature $T_h$ has a zero value at  $r_h=r_{dg}$. However, when $r_h< r_{dg}$, the horizon temperature becomes negative and forms an ultracold black hole. This result was discussed by Davies \cite{Davies:1978mf} who said that there are no obvious reasons from the thermodynamical viewpoint that prevent a BH temperature from becoming negative and linked this to a naked singularity. This is exactly what happened  in Fig. \ref{Fig:3}\subref{fig:3b} when $r_h<r_{min}$ region. The case of ultracold BH is explained by the existence of a phantom energy field \cite{Babichev:2014lda}, which investigates the decrease of the mass behavior in Fig. \ref{Fig:3}\subref{fig:3a}. When $r_h > r_{dg}$, the temperature becomes positive. When  $r_h$ becomes larger, the temperatures of both linear and nonlinear cases change in a similar manner.

Now  we are going to evaluate the heat capacity, $H_c$. Using  Eqs.~ (\ref{m55}), (\ref{m33}) and (\ref{m44}) we get:
\begin{eqnarray} \label{m66}
%&&{H_c}_{{}_{{}_{{}_{{}_{\tiny Eq. (\ref{sol1c})}}}}}=\frac{2\pi r_h{}^2\left[r_h{}^2-q^2\right]}{2(3q^2-2m r_h)}\,, \nonumber\\
%
 &&{ {H_c}_{{}_{{}_{{}_{{}_{\tiny Eq. (\ref{ass2n})}}}}}}=\frac{2\pi r_h{}^2\left[3\Lambda_{eff} r_h{}^4+r_h{}^2-q^2\right]}{2(\Lambda_{eff} r_h{}^4+3q^2-2m r_h)}\,.\nonumber\\
 &&\end{eqnarray}
 The above equation is not easy to get from it any information, thus we depicted it in Fig. \ref{Fig:3}\subref{fig:3c} with specific values of the parameters. As shown in  Fig. \ref{Fig:3}\subref{fig:3c}   that  both cases of linear and nonlinear charged BH solutions,   $H_c$ vanishes at   $r_{dg}$ and also their temperatures. In GR limit, the linear case, $H_c$ has positive  values when  $r_h>r_{dg}$, however,  when $r_h< r_{dg}$ it has negative values. In the NED case, the heat capacity is always positive unless $r_h<r_{dg}$.

\subsection{First law of thermodynamics of the BH solution (\ref{sol1cn})}\label{S61}

%%%%%%%%%%%%%%%%%%%%%%%%%%%%%%%%%%%%%%%%%%%%%%%%%%%%%%%%%%%%%%%%%%%%%%%%%%%%%%%%%%%%%%
Using Eq. (\ref{m33}) we get:
\begin{equation}\label{Mass1}
M={m_h}=\frac{\Lambda_{eff}r_h{}^3}{2}+\frac{r_h{}}{2}+\frac{q^2}{2r_h}\,.
\end{equation}
Moreover, from the definition of  entropy:
\begin{equation}\label{Mass}
S=\frac{A}{4}=\pi r_h{}^2\,,
\end{equation}
we can  show that the effective cosmological constant and  pressure are given as \cite{Wang:2020hjw}:
\begin{equation}
\label{1st}
P=\frac{3\Lambda_{eff}}{8\pi}\,.
\end{equation}
Eq. (\ref{Mass1}) can be rewritten in terms of pressure and entropy as:
\begin{equation}\label{Mass11}
M(S,q,P)=\frac{1}{6\sqrt{\pi S }}\left(3\pi\,q^2+3S+8P\, S^2\right)\,.
\end{equation}
 Therefore,  the parameters related  to $S$, $q$ and $P $ are calculated as:
\begin{eqnarray} \label{s1}
&&T=\left(\frac{\partial M}{\partial S}\right)_{P,q}=\frac{1}{4\pi r_h}\left(1-\frac{q^2}{r_h{}^2}+3\pi r_h{}^2\Lambda_{eff.}\right)\,, \nonumber\\
&&\xi=\left(\frac{\partial M}{\partial q}\right)_{S,P}=\frac{q}{r_h}\,, \qquad V=\left(\frac{\partial M}{\partial P}\right)_{S,q}=\frac{4}{3}\pi r_h{}^3\,,
\end{eqnarray}
with  $\xi$, T, and V are the  electric potential, temperature, and thermodynamic volume, respectively.  Using the above equations we can get the following Smarr relation
\begin{equation}\label{1st1}
M=2TS+\xi q-2VP\,,
\end{equation}
from which it is easy to prove  the
first law of thermodynamics as:
\begin{equation}\label{1st11}
dM=TdS+\xi\, dq+VdP\,.
\end{equation}
Equation (\ref{1st1}) ensures the validity of the first law of the BH (\ref{ass2n}).
\section{Discussion and conclusions}\label{S8}
In this research, we have considered spherically symmetric BH in Rastall's theory of gravity. We study the NED spherically symmetric spacetime and derive an exact solution that is affected by the Rastall parameter. This is the first time we derive a NED BH solution from the field equation of Rastall's gravitational theory. The main contribution of Rastall's parameter in this study comes from the contribution of the trace of the NED  which has a non-vanishing value in contrast to the linear Maxwell theory. We show that the effect of the Rastall parameter acts as a cosmological constant and the BH behaves asymptotically as (A)dS Reissner-Nordstr\"om spacetime. When the Rastall parameter vanishes, we get spacetime which asymptotes as flat Reissner-Nordstr\"om spacetime.

%Then we use the BH solution of the NED BH solution to extract its physical content. For this aim, we rewrite the metric potentials and show that in spite we begun with  unequal metric potential we got an analytic solution for the nonlinear charged case whose metric potentials are equal. In spite we have applied the field equations of  Rastall's theory without cosmological constant we have got an analytic BH solution with an effective cosmological solution. The only constrain on this effective cosmological constant is that the Rastall parameter must not equal to $-\frac{1}{4}$ as Eq. (\ref{ass2n}) showed.   We have shown that our BH solution of the nonlinear charged asymptotically behaves as (A)dS spacetime which means that the effect of the Rastall parameter in the frame of spherically symmetric solution behaves like a cosmological constant. This result is consistent with the study presented by Visser \cite{Visser:2017gpz}. This is one of the main results of the present study.

We have used the geodesic deviation to obtain the stability of the geodesic motion of the NED case. Furthermore, we investigated the horizons and demonstrated that the BHs presented in this study could have two horizons: the event horizon $r_1$, and the effective cosmological one $r_h$. Also, we fixed the minimum value of the BH mass that occurred at the degenerate horizon. We have also studied the thermal phase transitions and showed, in the linear electrodynamics case, i.e., $\epsilon=0$, the temperature became negative when $r_h < r_{d}$ and therefore, heat capacity became negative and thus we have unstable BH \cite{Chaloshtary:2019qvv,Sajadi:2019hzo,Yu:2019xdg,Ali:2019myr}. The same conclusions can be applied to the NED case. However, at $r_h>r_d$, we have a positive value of the $H_c$ which yields a stable BH. Finally, we proved the validity of the first law of thermodynamics. { It is worth noting that the result of thermodynamics presented in this study agrees with the study of thermodynamics presented in \cite{Caldarelli:1999xj} when the rotation parameter $a$ is vanishing.}

{{ In this study, we have discussed Rastall's theory using a special form of non-linear electrodynamics. This special form of non-linear electrodynamics reduces in our model to a linear form plus a cosmological constant.}   However, a deeper analysis is necessary, possibly regarding quantum effects in the universe. Meanwhile, the effects of Rastall's cosmology on the formation and properties of non-linear structures is a very promising research program. Furthermore, the study of $f(R)$-Rastall's theory will be extremely rich in the context of astrophysics \cite{Shahidi:2021lxt} . Within the frame of $f(R)$, a BH which is similar to Reissner-Nordstr\"om BH is presented \cite{Nashed:2019tuk} for a specific form of $f(R)$. Is it possible to derive a similar solution within Rastall's $f(R) $? This study will be carried out elsewhere.}

%\begin{acknowledgments}
%The author would like to thank the Referee for the constructive comments. Also, I would like to thank S. Nojiri for the useful discussion.
%\end{acknowledgments}

%%%%%%%%%%%%%%%%%%%%%%%%%%%%%%%%%%%%%%%%%%%%%%%%%%%%%%%%%%%%%%%%%%%%%%%%%%%%%%%%%%%%%%
%\bibliographystyle{apsrev}
%\bibliography{JRPHSRef}

\end{document}